\documentclass[11pt]{article}
\usepackage[margin=1in]{geometry}
\usepackage[utf8]{inputenc}
\usepackage[T1]{fontenc}
\usepackage{authblk}
\usepackage{hyperref}
\usepackage[normalem]{ulem}
\usepackage{graphicx}
\usepackage{multirow}
\usepackage{amsmath,amssymb,amsfonts}
\usepackage{amsthm}
\usepackage{mathrsfs}
\usepackage[title]{appendix}
\usepackage{xcolor}
\usepackage{textcomp}
\usepackage{manyfoot}
\usepackage{booktabs}
\usepackage{algorithm}
\usepackage{algorithmicx}
\usepackage{algpseudocode}
\usepackage{listings}
\usepackage{pdflscape}
\usepackage{rotating}
\usepackage{capt-of}
\usepackage{anyfontsize}
\usepackage{float}
\usepackage[numbers,sort&compress]{natbib}
\usepackage{subcaption}
\setcitestyle{square}
\DeclareRobustCommand{\secref}[1]{``\nameref{#1}''}
\providecommand{\keywords}[1]{\par\noindent\textbf{Keywords:} #1\par}
\providecommand{\bibcommenthead}{}
\providecommand{\arxivurl}[1]{\url{#1}}
\providecommand{\burl}[1]{\url{#1}}
\begin{document}

\title{Electric field control of anisotropic spin traffic between spin-filtering and magnon-scattering in a layered antiferromagnet}

% In your preamble:
% \usepackage[utf8]{inputenc} % if using pdfLaTeX
% \usepackage{authblk}

% In your preamble:
% \usepackage[utf8]{inputenc} % if using pdfLaTeX
% \usepackage{authblk}

% In your preamble:
% \usepackage[utf8]{inputenc} % if using pdfLaTeX
% \usepackage{authblk}

\author[1]{Mohamed Soliman\textsuperscript{\dag}}
\author[2,3]{Jaime Garrido Aldea\textsuperscript{\dag}}
\author[1]{Davy Borowski}
\author[2,4]{Dorye L. Esteras\textsuperscript{*}}
\author[1]{Neeraj. K. Rajak}
\author[5]{Niels Chapuis}
\author[5]{Abdelkarim Ouerghi}
\author[1]{Arnaud Gloppe}
\author[1]{J\'er\^ome Robert}
\author[1]{Bernard Doudin}
\author[6]{Takashi Taniguchi}
\author[6]{Kenji Watanabe}
\author[2,4]{Jos\'e H. Garc\'ia}
\author[2]{Stephan Roche}
\author[1]{Jean-Fran\c{c}ois Dayen\textsuperscript{*}}

\affil[1]{Institut de Physique et Chimie des Mat\'eriaux de Strasbourg (IPCMS), CNRS UMR 7504 \& Universit\'e de Strasbourg, 23 rue du Loess, 67034 Strasbourg, France}
\affil[2]{Catalan Institute of Nanoscience and Nanotechnology (ICN2), CSIC and The Barcelona Institute of Science and Technology (BIST), Campus UAB, 08193 Bellaterra, Spain}
\affil[3]{Universitat Aut\`onoma de Barcelona (UAB), 08193 Barcelona, Spain}
\affil[4]{Apeiron Intelligence, Start Up Lab Norte, 08018 Barcelona, Spain}
\affil[5]{Centre de Nanosciences et de Nanotechnologies (C2N), CNRS -- Universit\'e Paris-Saclay, 91120 Palaiseau, France}
\affil[6]{Research Center for Materials Nanoarchitectonics, National Institute for Materials Science, 1-1 Namiki, Tsukuba 305-0044, Japan}
\affil[ ]{\textsuperscript{\dag} These authors contributed equally}
\affil[ ]{\textsuperscript{*} Corresponding authors: dorye.esteras@icn2.cat; dayen@unistra.fr}


\begin{abstract}
Layered A-type antiferromagnetic (AFM) semiconductors offer a unique platform for spintronic devices by combining robust magnetism with gate-tunable semiconducting properties. We present a dual-gate field-effect transistor based on CrSBr, a layered A-type AFM with pronounced transport anisotropy and complex spin dynamics. Our device provides independent control over carrier density and displacement field. Systematic transport measurements reveal that the displacement field, rather than doping, predominantly governs the sign and magnitude of magnetoresistance ratio (MRR). We identify the origins of both negative and positive MRR components and connect them to interlayer and intralayer spin-dependent transport mechanisms. We interpret these findings with ab initio calculations and an effective spin-circuit model that reproduces the anisotropic magnetotransport. Our results identify the displacement field as an efficient electrical control parameter of spin transport in CrSBr and suggest that multi-gate architectures could provide similar control in other layered antiferromagnetic semiconductors.
\end{abstract}
\keywords{ 2D materials, magnetic semiconductors, antiferromagnetic semiconductors, negative magnetoresistance, magnetotransport}

\maketitle

\section{Introduction}

The emergence of quantum two-dimensional (2D) materials with spin-polarized band structures enables unprecedented control through electric and magnetic fields. This discovery marks a new paradigm in which spin, charge, and topology are intimately connected.  On one hand, the electric-field--induced breaking of spatial symmetries lifts the associated band degeneracies. On the other hand, the magnetic field manipulation of the spin polarization controls the time reversal symmetry in such systems. The combined effect opens the door to a broad spectrum of innovative property space \cite{gibertini2019magnetic, gong2019two, klein2018probing,lee2021magnetic,burch2018magnetism,wilson2021interlayer,huang2018electrical,zhang2019van,zhong2017van}.

This is particularly relevant for the recently discovered class of layered A-type antiferromagnetic semiconductors with intralayer ferromagnetic and interlayer antiferromagnetic orders. The interplay between applied electric fields, crystal symmetry, and time-reversal symmetry in these systems can produce highly tunable spin-polarized band structures. A recent example was reported by Fengrui Yao et al. in CrPS$_4$ where a dual-gate field-effect transistor architecture enabled displacement-field control of spin-current polarization in even-number-layer channels \cite{yao2025switching}. These findings extend and clarify the intriguing magnetotransport phenomena characteristic of this materials class.

Among the family of 2D A-type antiferromagnetic semiconductors, CrSBr stands out as a particularly compelling candidate. CrSBr's orthorhombic layers form quasi-one-dimensional Cr--S--Br chains along the b axis, a motif that governs both spin and charge responses. Magnetically, the crystal field and spin--orbit coupling impose a strong in-plane anisotropy with the easy axis parallel to the b axis, while weak interlayer exchange makes reorientation within the plane far easier than along the c axis. Transport mirrors the band structure: more dispersive bands and lighter carriers along the b axis enhance in-plane conductivity relative to the a axis, whereas the van der Waals gap enforces high resistivity along the c axis. Such tri-axial crystallographic electronic band structure and magnetic anisotropy give rise to a rich and tunable \textcolor{black}{magnetoresistance} response \cite{ziebel2024crsbr,Yang2021_triaxial,klein2023bulk}. Depending on temperature \cite{telford2020layered}, layer number \cite{ye2022layer}, and current orientation \cite{wu2022quasi}, CrSBr exhibits both positive and negative magnetoresistances (PMR and NMR, respectively), offering a versatile platform to probe the interplay between spin, charge, and lattice degrees of freedom.

These observations raise several questions about the microscopic origins of such intertwined features. For example, why does PMR appear only below 40 K in CrSBr, and why does it vanish when the magnetic field is parallel to the crystallographic easy axis? \cite{boix2022probing,telford2022coupling,cham2022anisotropic} What is the origin of the hidden order below 40 K? \cite{lopez2022dynamic} What is the relationship between band anisotropy and the current-direction dependence of \textcolor{black}{the magnetoresistance}? It has been shown that gating can modulate \textcolor{black}{magnetoresistance} in CrSBr using a single-gate field effect transistor (FET) architecture \cite{telford2022coupling,chou2025large,jo2024nonvolatile}. Is this modulation primarily driven by gate-induced doping (n) or by the displacement field (D), and how do these factors act in combination with band anisotropy? Moreover, clarifying how device architecture couples to microscopic transport pathways to set the spin-dependent current remains an open challenge. 

Addressing this fundamental understanding is pivotal to manipulate the spin current in these systems and engineer functional spintronics quantum devices. In this work, we combine ab initio calculations with tri-axial magnetotransport measurements of CrSBr bidirectional channels in a dual-gate FET architecture to investigate the microscopic origins of the anisotropic PMR and NMR components. The double-gate geometry enables disentangling the role of n and D in modulating \textcolor{black}{magnetoresistance}, revealing that D predominantly governs both its sign and magnitude. Calculations incorporating a vertical displacement field reveal band-structure splitting of layer and spin states, highlighting their key role in the gate control of magnetoresistance. Finally, we develop a spin circuit model that incorporates both the microscopic mechanisms and the device geometry. Within this model, we estimate how the spin-dependent current is distributed across the layers and how this distribution changes with the electric and magnetic fields. Together, these results make CrSBr a convenient platform to separate competing magnetotransport mechanisms in layered magnets and to study their electrical control.

\section{Results}

\subsection{Anisotropic magnetoresistance and magnon-electron interaction}\label{sec2}

\begin{figure}
    \centering
    \includegraphics[width=0.9\linewidth]{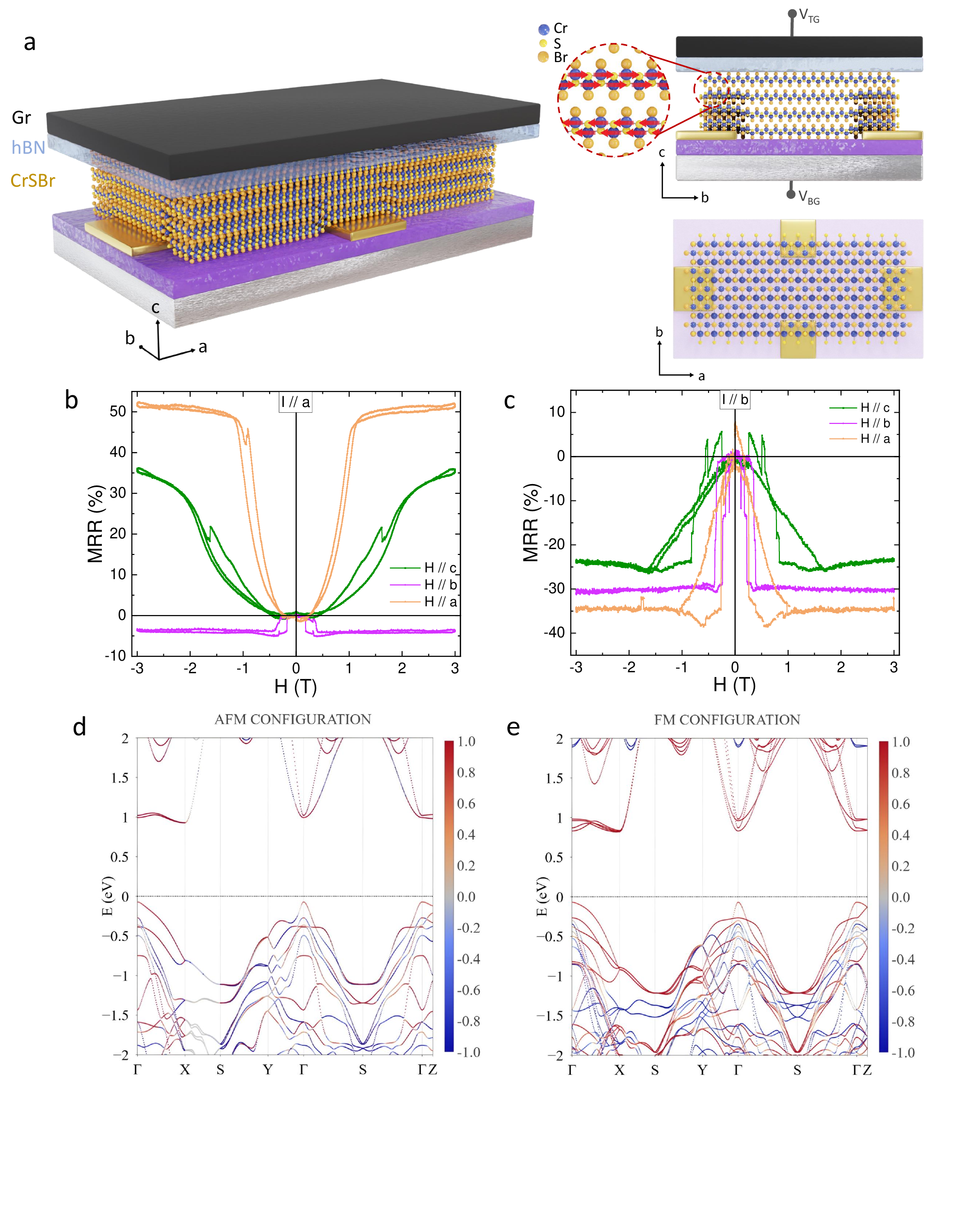}
    \caption{Dual gate anisotropic antiferromagnetic transistor. (a) (Left) Schematic of the CrSBr multi-axis dual-gate field-effect transistor, with current measured along the a and b axes of the channel. (Top-right) Figure shows the vertical device structure (arrows indicate crystallographic axes), while the magnified inset highlights the crystal structure along the easy magnetic b-axis, illustrating the antiferromagnetic texture. (Bottom-right) Top view of the electrode configuration relative to the CrSBr crystal. (b) Magnetoresistance ratio (MRR) as a function of magnetic field H applied parallel to the three crystallographic axes, with current along the a-axis, at $T = 10$~K. (c) Same as (b), with current along the b-axis. (d, e) Band structures of bulk CrSBr in the AFM (d) and FM (e) configurations obtained from DFT calculations. The colours measure the expected value of the $S_y$ operator for every state of the bands.}
        \label{PANEL1}
\end{figure}

\noindent To investigate the microscopic origin of magnetotransport in CrSBr, we designed a dual-gate FET platform. Figure~\ref{PANEL1}a illustrates a schematic of the device, which comprises multiple FETs fabricated around a CrSBr channel  (optical image of the device and the Atomic Force Microscopy (AFM) thickness traces of the CrSBr channel and the top gate dielectric hBN are provided in the supporting information Figure S.1). The architecture employs a Si/SiO$_2$ bottom gate, and a graphite top gate separated from the 30 nm CrSBr channel by an 8 nm hBN dielectric layer (see \nameref{sec7} for fabrication and setup details). The crystal structure of CrSBr is shown with the antiferromagnetic configuration (top-right schematic, Figure \ref{PANEL1}) and the orientation with respect to the device electrodes (Figure \ref{PANEL1}a bottom-right scheme). The quality of the magnetic properties and the crystal structure used in this work is confirmed by detailed superconducting quantum interference device (SQUID) measurements shown in the supplementary information, Figure \ref{fig:Slide1} and Figure \ref{fig:Slide2}. Furthermore, detailed Raman analysis is presented in the supplementary information, Figure \ref{fig:Raman}. Figures~\ref{PANEL1}b and \ref{PANEL1}c present the magnetoresistance ratio (MRR) as a function of magnetic field ($H$) for current applied along the $a$- and $b$-axes, respectively, where $\mathrm{MRR} = \big[(R - R_0)/R_0\big]\times100$ with $R_0 = R|_{H=0\,\mathrm{T}}$. For each current ($I$) direction, $H$ is applied sequentially along the three crystallographic axes ($a$, $b$, and $c$). For brevity, we denote the magnetoresistance ratio as $\mathrm{MRR}_{I,H}$, where $I \in \{a,b\}$ specifies the current direction and $H \in \{a,b,c\}$ the field orientation.

MRR of CrSBr exhibits a pronounced dependence on both \(I\) and \(H\) orientations. In Figure \ref{PANEL1}b when \(I\parallel a\) (the low-mobility axis), MRR is dominated by a large positive component (PMR) for \(H\parallel a\) (\(\mathrm{MRR}_{a,a}\) saturates at \(\approx50\%\)) and \(H\parallel c\) (\(\mathrm{MRR}_{a,c}\approx35\%\)), each superimposed on a weak negative contribution at low fields. In stark contrast, for \(H\parallel b\) (the magnetic easy axis), only negative MRR (NMR) emerges, saturating at \(\approx -5\%\). The disappearance of PMR when $H \parallel b$ points to an origin in CrSBr's intrinsic magnetic anisotropy.
Figure \ref{PANEL1}c shows \(I\parallel b\) (the high-mobility axis); MRR is exclusively negative (\(-25\%\) to \(-35\%\)) for all three \(H\) orientations, with saturation fields identical to those for \(I\parallel a\) (Fig. \ref{PANEL1}b). Remarkably, in the case of the absence of PMR (when \(H\parallel b\)), the NMR magnitude is greater for \(I\parallel b\) than for \(I\parallel a\). We also observe a memory effect following ferromagnetic-like switching. This is further confirmed by performing successive back-and-forth sweeping within the same field polarity range, and observing the disappearance of the switching in the second cycle. Similar features have been previously observed in references \cite{telford2022coupling} and \cite{ye2022layer}. Thanks to the weak interlayer antiferromagnetic coupling of CrSBr, the switching of the layers is expected to be hysteretic \cite{wang2026ferromagnet}. Besides, the possible origin could be associated to defects-induced domain wall pinning \cite{ye2022layer,telford2022coupling}, or the existence of frustrated or disordered magnetic states with no long-range order \cite{lopez2022dynamic}. It is worth mentioning that the noise level observed on the MRR traces for
\(I\parallel b\) direction is around \(0.2\%\) while it is around \(0.1\%\) for the \(I\parallel a\) direction. This could be related to the difference in electrode width (500 nm for the b direction electrodes and 5 $\mu$m for the a direction).

To elucidate the origin of the NMR, we perform density functional theory (DFT) calculations (see \nameref{sec7} and the Supporting Information section \secref{supplementary section DFT Bands}). Figure~\ref{PANEL1}d-e shows the bulk CrSBr band structures in AFM and FM interlayer configurations, where the stronger dispersion along $\Gamma$--Y compared to $\Gamma$--X reflects the anisotropic conductivity. By contrast, the extremely flat $\Gamma$--Z band (momentum along the $c$-axis) highlights the weak interlayer coupling. The narrow $\Gamma$--Z bandwidth is proportional to the orbital overlap between adjacent layers, indicating a very small interlayer tunneling probability.

We attribute the dominant NMR observed for $I \parallel b$ to the combined influence of band-structure anisotropy and interlayer spin-dependent tunneling (see the Supporting Information section \secref{ssec:results_circuit} and fig. \ref{fig:result_circuit_rho}). Within this framework, the NMR is enhanced when conduction electrons extend across the full CrSBr channel thickness, and reduced when transport is confined to a few layers near the injection electrodes.

The general attribution of NMR in multilayer CrSBr to the interlayer spin filtering effect is consistent with previous studies \cite{telford2020layered,boix2022probing}, with negligible contribution from the individual layers previously hypothesized to originate from interaction with defects  \cite{telford2022coupling}. On the contrary, the PMR is believed to be an intralayer contribution with a debated physical origin.  

\begin{figure}[H]
    \centering
    \includegraphics[width=1\linewidth]{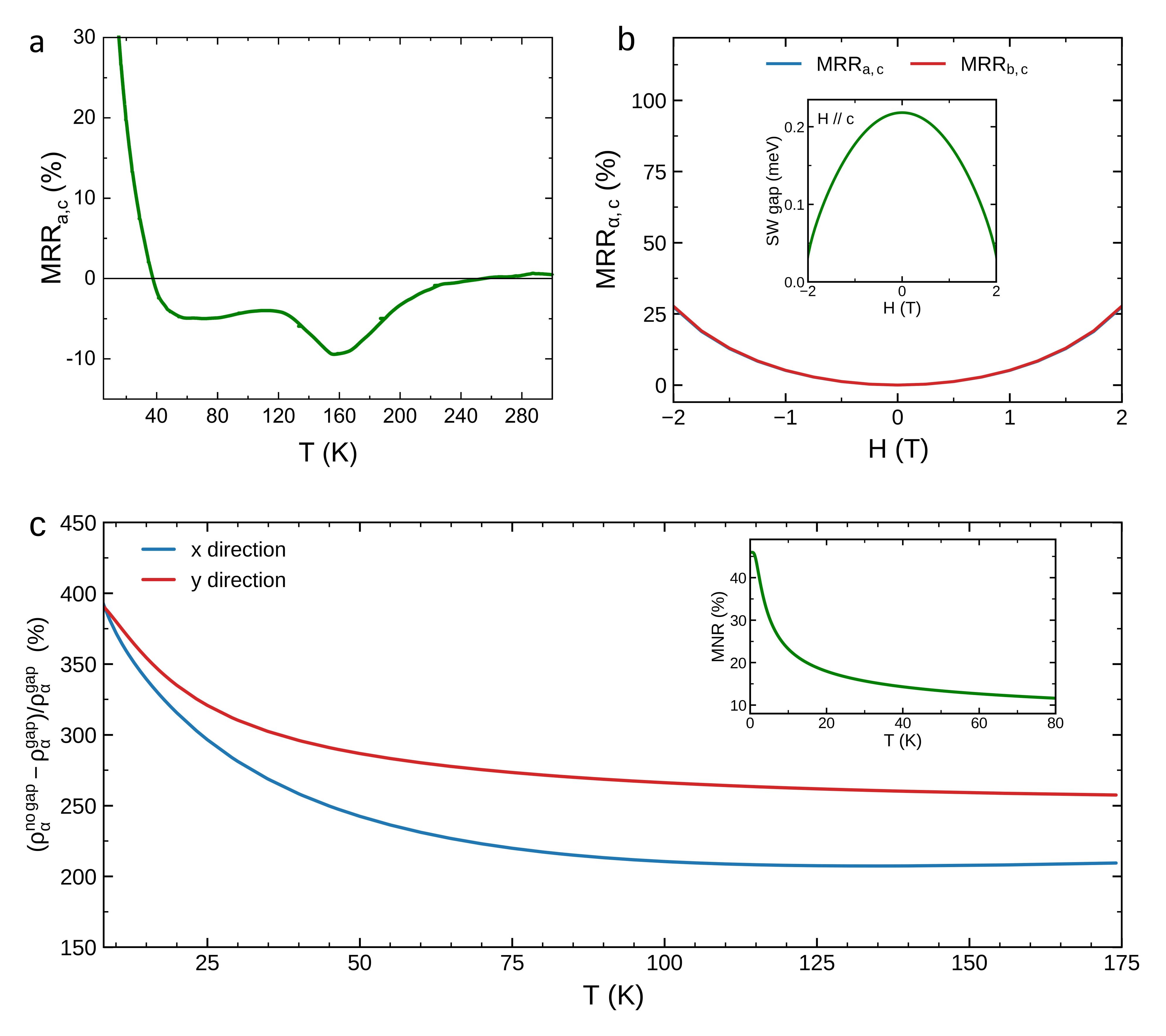}
    \caption{Magnon assisted transport and spin-wave gap modulation. (a) MRR$_{a,c}$ at H = -3T vs T. (b) Calculated magnon-electron MRR$_{I,c}$ for $I \parallel a$ and $I \parallel b$ (the two curves are superimposed). (Inset b)  Evolution of the calculated spin-wave gap when the magnetic field is applied along the c-axis for T = 0 K. (c) Behaviour with temperature of $\frac{\rho_\alpha^{ \text{nogap} } - \rho_\alpha^{ \text{gap} } }{\rho_\alpha^{ \text{gap} }}$ (\%), where $\rho_\alpha$ is the calculated $\alpha$-directional magnon-electron resistivity with or without a gapped magnon spectrum. (Inset c) The magnon number ratio (MNR) as defined in equation \eqref{MNR equation} vs temperature.}
    \label{Figures_CrSBr_2}
\end{figure}

Interestingly, the PMR component has been reported only below 40 K \cite{telford2020layered,boix2022probing,telford2022coupling}, consistent with our observations in Figure~\ref{Figures_CrSBr_2}a, which shows the temperature dependence of MRR$_{a,c}$. At 40 K, the MRR changes sign to NMR, which develops into a shallow parabola around 160 K, the Curie temperature of monolayer CrSBr. This NMR feature arises from the magnetic-field suppression of spin fluctuations near the ferromagnetic--paramagnetic phase transition, a hallmark of magnon--electron scattering \cite{Yamada1972,Yamada1973}.

To identify the origin of the PMR component, we considered two key observations: its dependence on the crystallographic magnetic anisotropy (Figure~\ref{PANEL1}b) and the influence of magnon--electron scattering on the resistance in CrSBr. To rationalize these findings,  we combined our DFT calculations below with spin-wave calculations (Supporting Information sections \secref{section TB2J}, \secref{section magnonic properties} and \secref{sec:magnon_electron_model}), the field-dependent magnon population (\secref{section magnonic properties}) and a transport-weighted magnon--electron scattering treatment (\secref{sec:magnon_electron_model}, \secref{sec:implementation} and \secref{sec:results}). The Heisenberg Hamiltonian (Eq.~\eqref{H}) model allowed us to extract the spin-wave dispersions and thus establish a microscopic framework for the role of anisotropy and spin-wave excitations (\nameref{sec7}; Supporting Information sections \secref{section TB2J}, \secref{section magnonic properties} and \secref{sec:magnon_electron_model}). We further extended the magnetic Hamiltonian by introducing a Zeeman-like term \cite{Evans2014} to account for an external magnetic field and track the evolution of the magnon gap (inset figure~\ref{Figures_CrSBr_2}b):

\begin{equation}
    H = \sum_{i,j} \mathbf{S}_i \cdot \mathbf{J}_{ij} \cdot \mathbf{S}_j 
      + \sum_{i} \mathbf{S}_i \cdot \mathbf{A}_{i} \cdot \mathbf{S}_i 
      - \mathbf{H} \cdot \sum_i \mathbf{S}_i ,
    \label{H}
\end{equation}

where $\mathbf{J}_{ij}$ is the exchange tensor including both isotropic and anisotropic interactions, 
$\mathbf{A}_i$ is the single-ion anisotropy (SIA) tensor, 
and $\mathbf{H}$ is the external magnetic field vector.

The spin-wave spectra and the associated gap were calculated including the additional Zeeman-like term for several values of the magnetic field (Supporting Information section \secref{section magnonic properties}). The resulting gap evolution, shown in the inset of Figure~\ref{Figures_CrSBr_2}b, confirms that applying a field perpendicular to the easy axis reduces the spin-wave gap, thereby enhancing magnon population as calculated in figure \ref{fig:result_Nmag_rho_B} left (\secref{section magnonic properties}). This magnon population enhancement increases the probability of magnon-electron scattering, a scattering mechanism that was observed in Figure \ref{Figures_CrSBr_2}a as previously mentioned.

At any given temperature, this resistivity is influenced by two occupation factors entering the scattering rate (Supporting Information sections \secref{ssec:FGR_transverse} and \secref{ssec:FGR_longitudinal}, and eqs. \ref{eq:tau_trans_impl} and \ref{eq:tau_long_impl}): the number of magnons available to scatter from (governed by the Bose-Einstein distribution) and the number of accessible electronic final states (governed by the Fermi-Dirac distribution) (\secref{sec:magnon_electron_model} and \secref{sec:implementation}). The interplay of these two factors defines the dynamics of the magnon-electron resistivity with temperature and magnetic field. Consequently, the calculated magnon-electron resistivities grow as the spinwave gap closes (fig. \ref{Figures_CrSBr_2}b; \secref{sec:magnon_electron_model}, \secref{sec:implementation} and \secref{sec:results}).  

Figure \ref{Figures_CrSBr_2}b shows the magnon-electron MRR$_{\alpha,c}$ calculated from the magnon-induced resistivity vs H. The MRR crosses 25$\%$ at saturation (the calculation accuracy drops after saturation; fig. \ref{fig:imag_energies} and Supporting Information section \secref{ssec:impl_LSWT_validity}). Similar results are obtained for MRR$_{\alpha,\text{a}}$ (Fig. \ref{fig:result_MR_B_Hpa} and section \secref{ssec:results_Hpa}). Interestingly, the calculated magnon MRR is identical for both current directions despite the anisotropic conduction nature of CrSBr.

To quantify the evolution of the total magnon population with temperature ($T$), the inset of figure~\ref{Figures_CrSBr_2}c shows the magnon number ratio (MNR) between the gapped and the ungapped spectrum as a function of $T$:
\begin{equation}
\mathrm{MNR} = \frac{N_{\text{magnon}}^{\text{no gap}} - N_{\text{magnon}}^{\text{gap}}}{N_{\text{magnon}}^{\text{no gap}}} \times 100 ,
\label{MNR equation}
\end{equation}
where $N_{\text{magnon}}^{\text{no gap}}$ and $N_{\text{magnon}}^{\text{gap}}$ denote the magnon numbers without and with magnon gap respectively (see the Supporting Information section \secref{section magnonic properties} for details of the calculation).

The MNR decreases with temperature and tends to saturate above approximately 40 K. Interestingly, the temperature evolution of the magnon-electron resistivity upon magnon gap removal follows the same trend. Figure \ref{Figures_CrSBr_2}c plots the relative change $\frac{\rho_\alpha^{ \text{nogap} } - \rho_\alpha^{ \text{gap} } }{\rho_\alpha^{ \text{gap} }}$ (\%), where $\rho_\alpha$ is the $\alpha$-directional magnon-electron resistivity with or without gapped magnon (Supporting Information sections \secref{sec:magnon_electron_model}, \secref{sec:implementation} and \secref{sec:results}). The ratio decreases from low temperatures and above $\sim$40 K; it saturates for both current directions.

At low temperature, closing the spin-wave gap with a hard-axis magnetic field produces a large relative increase in low-energy magnetic excitations, enhancing magnon--electron scattering and giving rise to observed PMR. At higher temperature, thermal magnons are already populated at zero field, so the additional field-induced change in the magnon population becomes comparatively smaller. The PMR contribution therefore weakens and eventually becomes masked by the NMR background above approximately 40 K. For the low mobility axes, the drop in this ratio is more prominent than for the high mobility axes.

In summary, we propose the magnon scattering magnetic field modulation as the mechanism responsible for the CrSBr layer PMR. When H is applied along any of the two hard axes, the competition with the crystallographic anisotropic field closes the gap in the magnon energy spectrum. At low T, the gap closure yields population of the low-energy magnonic states that are responsible for the majority of the magnon-electron scattering events. This H-induced increase in the magnon-electron scattering rate increases the resistivity, resulting in the measured PMR. Note that this naturally explains the absence of PMR when H is applied along the easy axis (in this case, H is not competing with the crystallographic anisotropic field beyond the spin-flip transition). Moreover, the weakening of the PMR with T is also captured, as at high T the low-energy magnonic states are thermally populated, and the H effect on magnon-electron scattering is reduced. Our theoretical calculations quantitatively confirm this understanding and successfully reproduce the trends with magnetic field orientations and the evolution with T. Further experimental and theoretical investigations are still needed to quantify the relative strength of the magnon-electron scattering mechanism to the other scattering mechanisms in play. Studying magnon-induced MRR in magnetic semiconductors is an interesting yet uncharted territory. Our work motivates further exploration of this research area with promising results.

\subsection{Doping versus Displacement field effects on MRR}\label{sec4} 

\begin{figure}[H]
    \centering
    \includegraphics[width=1\linewidth]{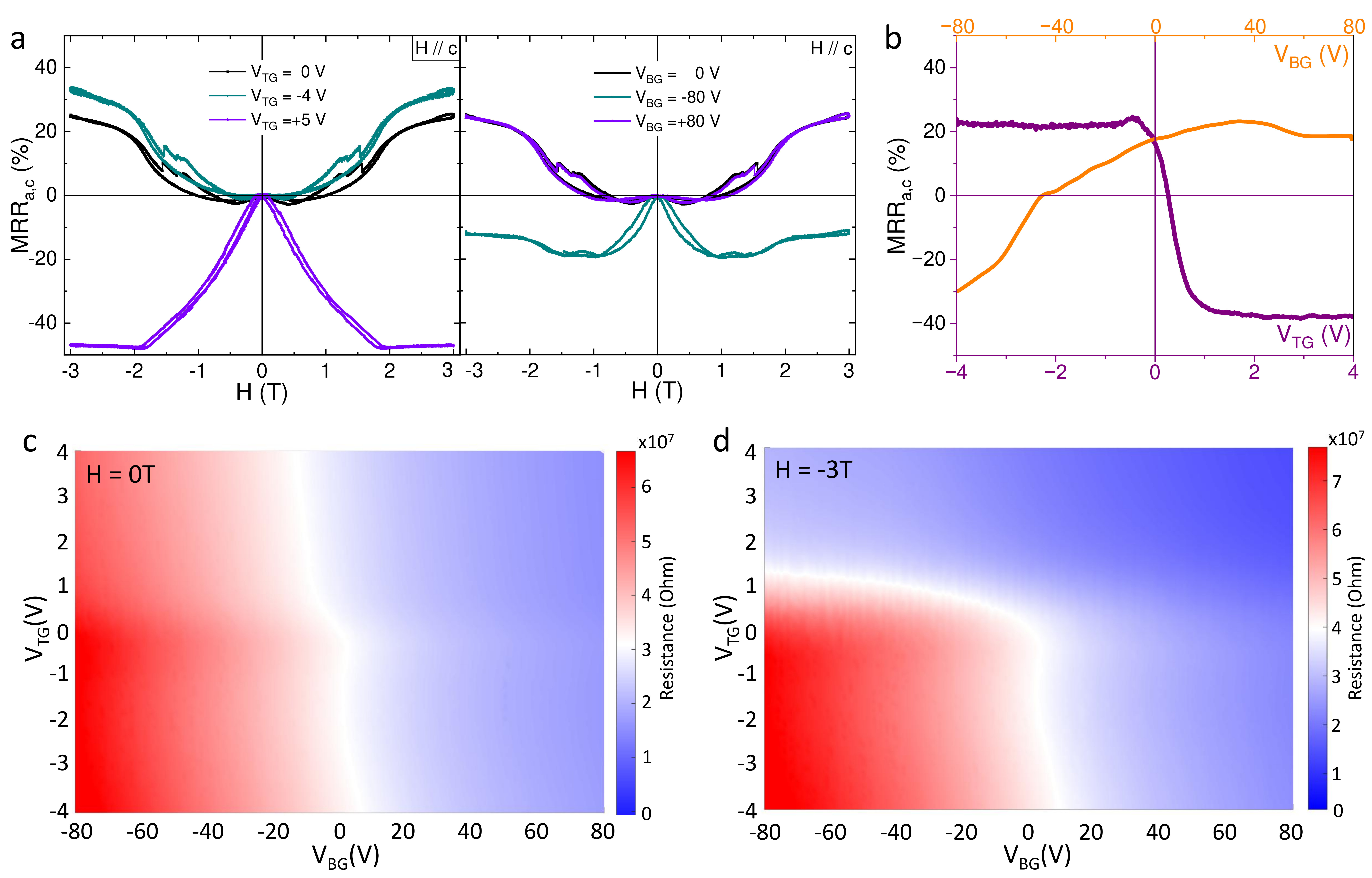}
    \caption{Gate-resolved MRR. 
(a) \(\mathrm{MRR}_{a,c}\) vs.\ \(H\) for top-gate (\(\sim 8\,\mathrm{nm}\) hBN dielectric) voltages \(V_{\mathrm{TG}} = 0, -4, +5~\mathrm{V}\) (left graph). 
\(\mathrm{MRR}_{a,c}\) vs.\ \(H\) for bottom-gate (\(285\,\mathrm{nm}\) \(\mathrm{SiO_2}\) dielectric) voltages \(V_{\mathrm{BG}} = 0, -80, +80~\mathrm{V}\) (right graph). 
(b) \(\mathrm{MRR}_{a,c}\) traces (extracted at \(H=-3~\mathrm{T}\)) vs.\ \(V_{\mathrm{TG}}\) (purple color) and \(V_{\mathrm{BG}}\) (orange color). 
(c) Color map of the resistance (current along the \(a\)-axis) as a function of \(V_{\mathrm{TG}}\) and \(V_{\mathrm{BG}}\) at \(H=0\) in the antiferromagnetic state. 
(d) Color map of the resistance (current along the \(a\)-axis) as a function of \(V_{\mathrm{TG}}\) and \(V_{\mathrm{BG}}\) at \(H=-3~\mathrm{T}\) in the induced ferromagnetic state. T = 10K} 
    \label{MRR with electric field}
\end{figure}

We now examine the effect of electrostatic gating on MRR by independently varying each gate in the dual-gate architecture. In Figure \ref{MRR with electric field}a, we investigate the influence of the top gate bias on \(\mathrm{MRR}_{a,c}\)  while the bottom gate is held at zero voltage (left graph). A positive \(V_{\mathrm{TG}}\) drives \(\mathrm{MRR}_{a,c}\) from positive to negative, whereas a negative \(V_{\mathrm{TG}}\) enhances the PMR. In contrast, Figure \ref{MRR with electric field}a (right graph) shows that the bottom-gate voltage \(V_{\mathrm{BG}}\) has an asymmetric influence: a positive \(V_{\mathrm{BG}}\) has little effect on \(\mathrm{MRR}_{a,c}\), while a negative \(V_{\mathrm{BG}}\) flips \(\mathrm{MRR}_{a,c}\) from positive to negative. The dependence of the saturated \(\mathrm{MRR}_{a,c}\) amplitude (at -3 T) on each gate voltage is confirmed in Figure \ref{MRR with electric field}b showing \(\mathrm{MRR}_{a,c}\)(\(V_{\mathrm{BG}}\)) and \(\mathrm{MRR}_{a,c}\)(\(V_{\mathrm{TG}}\)). The top-gate trace crosses zero near \(+0.5\)\,V, whereas the bottom-gate trace, with opposite gate dependency, crosses zero near  \(-55\)\,V. The opposite slopes of the top and bottom gate dependency pinpoint the vertical displacement field (\(D\)) across the CrSBr layers as the driving parameter for the MRR sign change. 

To disentangle the roles of the displacement field (D) and doping (n) in our device, we carefully investigated the interplay between the electrostatic coupling of the gates to the CrSBr channel and its magnetic states. Figure \ref{MRR with electric field}c and Figure \ref{MRR with electric field}d show the colormaps of the channel resistance (R) as a function of the two gate voltages for (H = 0~$\mathrm{T}$) and (H = -3~$\mathrm{T}$), respectively. Remarkably, at (H = 0~$\mathrm{T}$), the bottom gate emerges as the primary driver of the FET resistance modulation, with a negligible contribution from the top gate. This behaviour can be understood by considering the bottom-contact configuration together with the antiferromagnetic spin-filtering effect in the CrSBr channel. The combination of these two effects prevents the top gate from efficiently accumulating charge at the top interface of the CrSBr channel because of the high vertical resistance imposed by spin filtering. The ability of the top gate to effectively dope the channel increases dramatically when the spin-filtering condition is lifted at (-3~$\mathrm{T}$), as shown in Figure \ref{MRR with electric field}d.

\begin{figure}[H]
    \centering
    \includegraphics[width=1\linewidth]{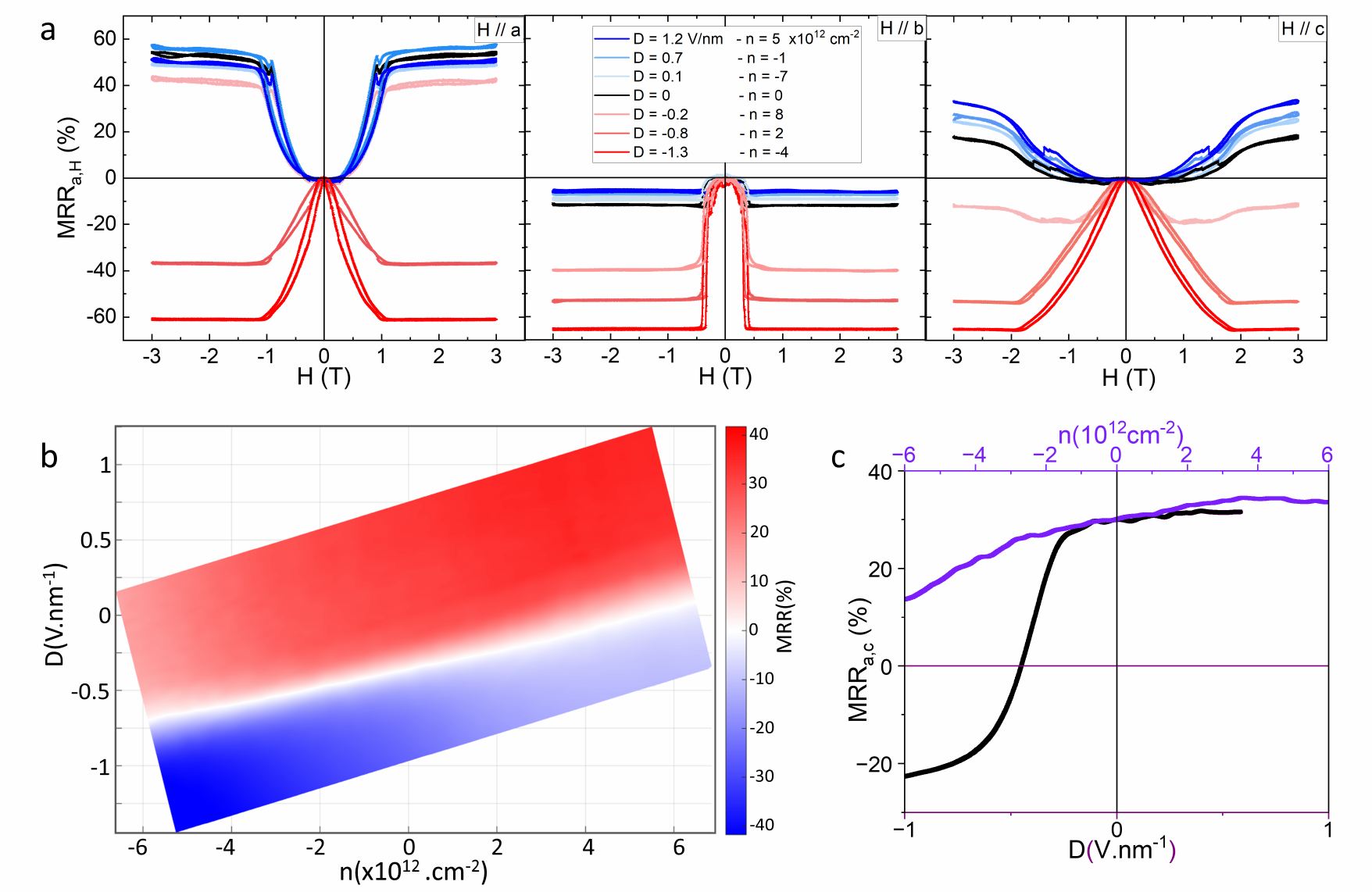}
    \caption{Disentangling the role of D and n on the Magnetoresistance control. (a) \(\mathrm{MRR}_{a,H}\) vs H, for H // a-axis (top), H // b-axis (middle), and H // c-axis (bottom). The traces represent different (D,n) combinations maximizing either D or n, with complementary intermediate cases. (b) 
    $\mathrm{MRR}_{a,c}(D,n)$ color map showing the variation of the magnetoresistance \(\mathrm{MRR}_{a,c}\) along the a axis for H // c-axis, as a function of the electrostatic gate-controlled carrier density n, and the displacement field D. (c) Extracted line cuts of $\mathrm{MRR}(D)$ at $n=0$ and $\mathrm{MRR}(n)$ at $D=0$.}
    \label{MRR with n and D}
\end{figure}

Figure~\ref{MRR with n and D} examines the gate dependence of the MRR for a set of gate configurations selected to maximize either \(D\) or \(n\), together with several intermediate values. For each trace, we calculated the corresponding \(D\) and \(n\). The legend is arranged in ascending order of the MRR value at \(-3~\mathrm{T}\). Consistent with the previous observations, the \(D\) values follow the trend of the saturated MRR amplitude, whereas the \(n\) values show no clear correlation. Remarkably, when \(I \parallel a\), this behaviour is observed for all three magnetic-field orientations (\(a\), \(b\), and \(c\)). For \(\mathrm{MRR}_{a,a}\) (Figure~\ref{MRR with n and D}a-top), \(+D\) has little effect, whereas increasing \(-D\) progressively enhances the NMR component, ultimately reversing the MRR completely from \(\sim+60\%\) to \(\sim-60\%\). Similarly, for \(\mathrm{MRR}_{a,b}\) (Figure~\ref{MRR with n and D}a-middle) and \(\mathrm{MRR}_{a,c}\) (Figure~\ref{MRR with n and D}a-bottom), \(+D\) suppresses the NMR component, whereas \(-D\) amplifies it. At the maximum \(-D\), all field orientations yield \(\mathrm{MRR}_{a,H}\approx -60\%\). At the maximum \(+D\), the saturated PMR reaches \(\sim+60\%\) for \(H\parallel a\) and \(\sim+35\%\) for \(H\parallel c\). By contrast, for \(H\parallel b\), the PMR remains absent, and the NMR is minimized (\(\sim3\%\)), consistent with the magnon interpretation. These trends confirm that the PMR strength depends on the magnetic-field direction, in contrast to the more isotropic NMR. On the other hand, when \(I\) flows along the high-mobility \(b\)-axis, \(D\) has no clear effect on the MRR (see Supplementary Information, figure \ref{fig:MRR_n}).

To further disentangle the effects of $D$ and $n$, we first calculated $\mathrm{MRR}(V_{\mathrm{TG}},V_{\mathrm{BG}})$ from the resistance colormaps (Figure~\ref{MRR with electric field}), and then applied a coordinate transformation to obtain $\mathrm{MRR}(D,n)$, shown in Figure~\ref{MRR with n and D}b. The displacement field $D$ is defined in the same way for both the antiferromagnetic (AFM) state and the field-induced ferromagnetic (FM) state (see the Supporting Information section \secref{supplementary: doping section}). In contrast, the carrier density $n$ must be evaluated separately for the AF and iFM states because the top gate affects the doping differently in the two cases, as discussed above. These two definitions correspond to the two limiting values of $n$ relevant to the MRR (details of the coordinate transformation for both cases are provided in the Supporting Information section \secref{supplementary: doping section}). Importantly, both approaches lead to the same overall trends and physical interpretation, as discussed below. For clarity, we present here the case where $n$ is calculated for the AFM state (Figure~\ref{MRR with n and D}b), while the corresponding FM result is shown in Supporting Information Figure~\ref{fig:MRR_nD_3T}. This transformation allows for plotting the $\mathrm{MRR}(D,n)$ colormap along with extracted line cuts of $\mathrm{MRR}(D)$ at $n=0$ and $\mathrm{MRR}(n)$ at $D=0$ (Figure~\ref{MRR with n and D}c). Together, these data demonstrate that $D$ is the primary parameter governing the MRR, as it strongly affects both its amplitude and its sign reversal. In comparison, $n$ induces only a relatively minor modulation of the MRR.

\begin{figure}[H]
    \centering
    \includegraphics[width=1\linewidth]{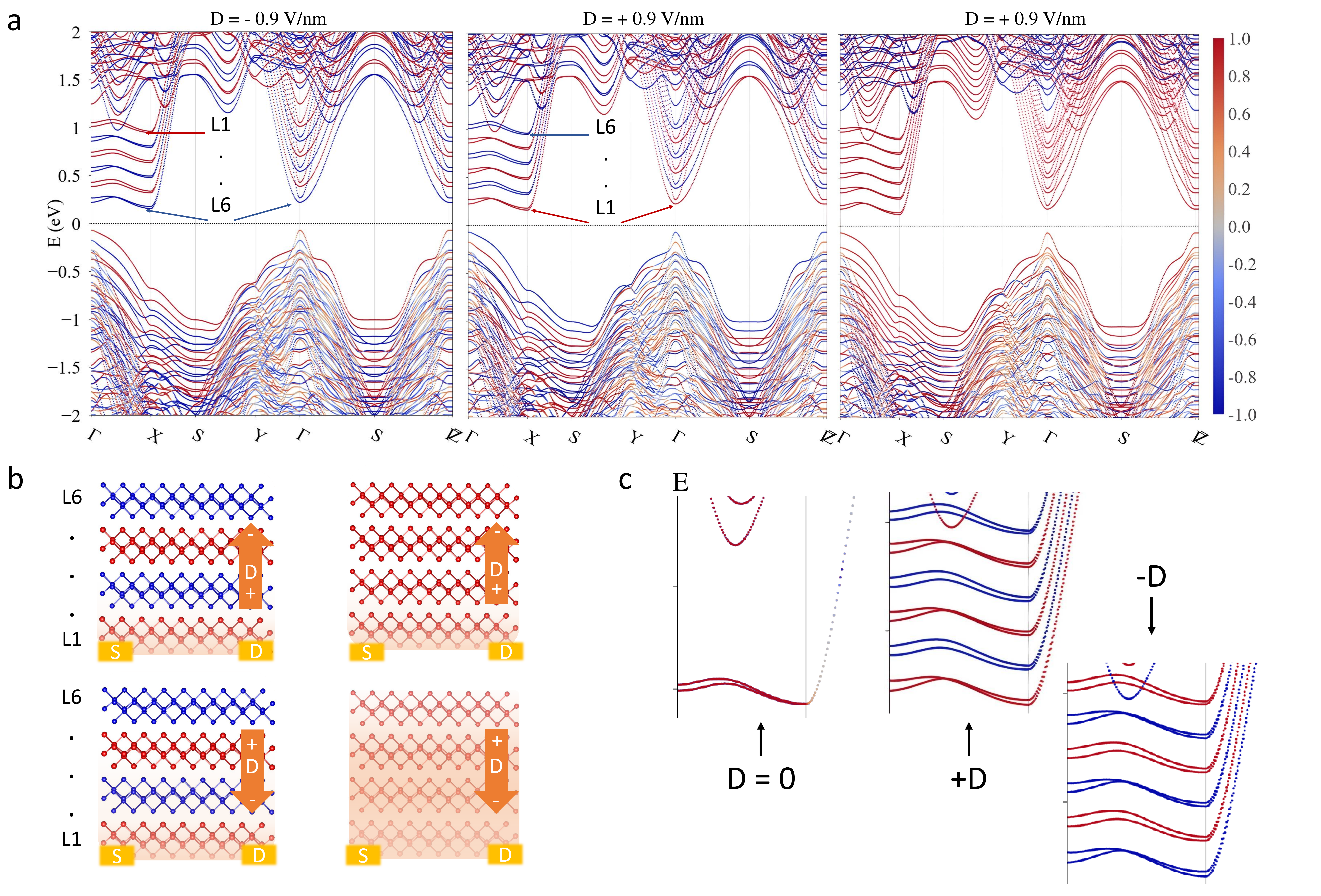}
    \caption{Displacement field control of spin-polarized bands. (a) Spin-resolved band structures of 6-layer CrSBr in the AFM (left and center) and FM configurations (right), shown at opposite displacement fields. The colours measure the expected value of the $S_y$ operator for every state of the bands. (b,c) Schematics illustrating the effect of the electric field on charge distribution among the layers in our bottom-contacted device (b) and on the spin-polarized band structure (c).}
    \label{Figures_CrSBr_4_new}
\end{figure}

The effect of the displacement field $D$ on the CrSBr band structure is shown in Figure~\ref{Figures_CrSBr_4_new}. Figure~\ref{Figures_CrSBr_4_new}a displays the AFM state of a six-layer CrSBr cell at $D=-0.9~$ V/nm (left), and $D=0.9~$ V/nm (middle). Applying $D$ along the $c$-axis lifts both spin and layer degeneracies. The resulting Stark shift forms a potential staircase across the layers, with alternating spin flavours dictated by the AFM order. In the FM state, the Stark shift is identical, but all layers carry the same spin character (Figure~\ref{Figures_CrSBr_4_new}a right). Positive and negative $D$ produce symmetric band diagrams; however, tracking the individual layer levels reveals that reversing $D$ reverses the direction of the staircase (Figure~\ref{Figures_CrSBr_4_new}a and figures \ref{AFM electric field} and \ref{FM electric field}).

Because of the device geometry (bottom contacts), electron injection occurs at the bottom layer (layer 1 in our convention). From there, electrons may diffuse into higher layers provided energy and momentum conservation rules are satisfied. For the nominal field used in the DFT calculation, the layer-dependent Stark splitting becomes comparable to, or larger than, the narrow ($\Gamma$-X) bandwidth, indicating that even a moderate layer-resolved potential gradient can strongly affect interlayer accessibility for carriers moving along the a-axis. For $I \parallel a$, a positive $D$ places layer 1 at the bottom of the ladder, confining $\Gamma$--X electrons to the injection layer even after spin filtering is lifted. Conversely, a negative $D$ puts layer 1 at the top of the ladder, favouring interlayer tunnelling through successive layers (Figure \ref{Figures_CrSBr_4_new}c). Upon spin alignment by an external $H$, the filtering condition is removed, giving rise to a pronounced NMR. Along the $b$-axis, however, the wide $\Gamma$--Y band spans the entire Stark-split window. Thus, for $I \parallel b$, a continuum of states (up to $eV_\text{D}$) enables interlayer tunnelling (by changing $k$-state) without energy penalty (Figure \ref{Figures_CrSBr_4_new}c). Even under $+D$, tunnelling remains efficient once spin filtering is lifted, producing a strong NMR largely unaffected by $D$.

In summary, when current flows along the low-mobility $a$-axis, a negative $D$ enhances NMR by delocalizing conduction across multiple layers, whereas a positive $D$ suppresses NMR by confining transport to the injection layer (illustration in Figure \ref{Figures_CrSBr_4_new}b left). In contrast, along the high-mobility $b$-axis, the broad conduction band eliminates the energy penalty for interlayer tunnelling, rendering the NMR largely insensitive to $D$.

\subsection{Decoupling spin-filtering and magnon scattering interactions with circuit modelling}\label{sec5}

The NMR and PMR mechanisms coexist in our device, and the measured MRR reflects
their combined effects. Our proposed picture is that the NMR originates from the reduction of interlayer resistance ($R_{\mathrm{inter}}$) under a magnetic field, driven by spin-filtering relaxation. In contrast, the PMR arises from the increase in intralayer resistance ($R_{\mathrm{intra}}$), caused by magnon-electron scattering within the layers. Our magnon-electron transport calculations show that this magnon-scattering contribution is strongly anisotropic, with the magnon-electron layer resistivity about $19\times$ larger along $a$ than along $b$ (Figure \ref{fig:result_sigma_ratio_B}). Feeding this computed layer resistivity into a minimal ladder circuit, with the interlayer term described by the tunnelling-magnetoresistance equation, shows that this single anisotropy is enough to produce an opposite magnetoresistance sign along the two directions (Figs.~\ref{fig:result_circuit_rho} and \ref{fig:result_circuit_MR}): along $a$ the large layer resistivity dominates and the response tracks the positive magnon term, while along $b$ the small layer resistivity lets the interlayer TMR term control the field dependence, giving a negative
magnetoresistance. This directional separation of the two mechanisms motivates a
circuit model of the CrSBr channel, represented as a network of variable
resistances, which enables a quantitative separation of the spin-filtering and
magnon-scattering contributions to the measured MRR under different displacement
fields ($D$). A qualitative motivation for this description, together with the
sign-reproducing ladder calculation, is provided in the Supplementary Information
(section \secref{ssec:results_circuit} and Figs.~\ref{fig:result_circuit_rho} and \ref{fig:result_circuit_MR}).

\begin{figure}[H]
    \centering
    \includegraphics[width=1\linewidth]{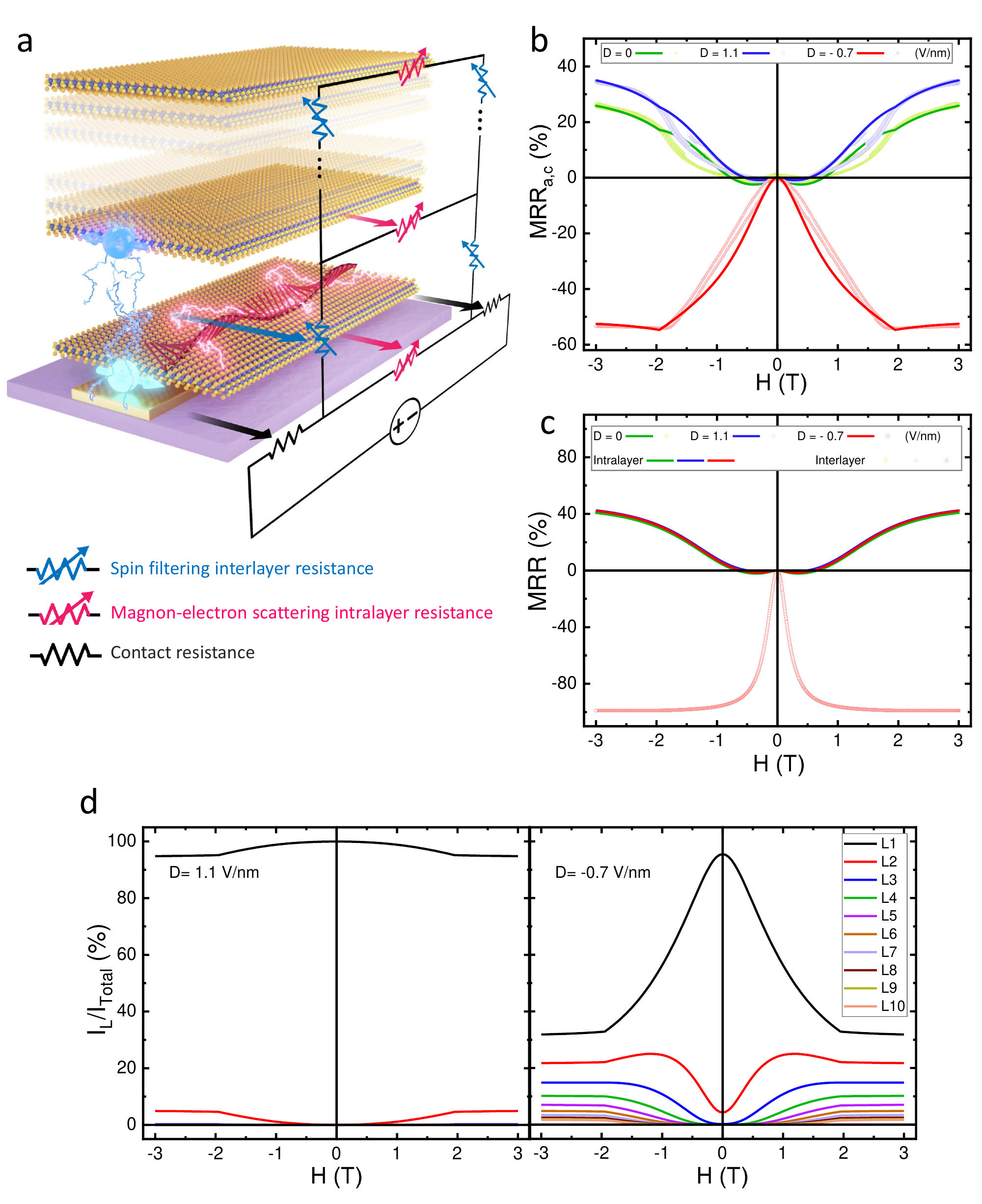}
    \caption{Spin circuit model and layer-resolved spin transport. (a) Equivalent circuit model of the device and scheme of its corresponding device part. Red resistors represent magnon-electron scattering intralayer resistances governing the PMR contribution, while blue resistors represent interlayer resistances controlling the NMR. Black resistors are contact resistances. (b) Fit of the experimental MRR with the circuit model for three different values of the displacement field. (c) Intralayer and interlayer MRR for three different values of the displacement field using eqs.~\eqref{Interlayer resistance} and \eqref{Rintra}. (d) Current normalized to the total current passing through each layer for the positive (left graph) and negative (right graph) displacement field cases. These results correspond to the case in which the current flows along the $a$-axis and the magnetic field is applied along the $c$-axis. The fits were obtained as shown in the Supporting Information section \secref{fitting PMR supplementary section} with the total number of magnons at $T$ = 10 K.}

    \label{Figures_CrSBr_6}
\end{figure}

Figure~\ref{Figures_CrSBr_6}a illustrates the equivalent circuit of the CrSBr channel. The channel is modeled as a ladder network consisting of ten layers, each layer comprising an intralayer resistance $R_{\mathrm{intra}}$ connected to adjacent layers via interlayer resistances $R_{\mathrm{inter}}$. Source and drain contacts are represented by contact resistances ($R_{c}$), both connected to the bottom layer in accordance with the device geometry. All layers $R_{\mathrm{intra}}$ and $R_{\mathrm{inter}}$ are assumed equal, given the negligible effect of doping.

\(R_{\mathrm{inter}}\), models the spin‐filtering mechanism. Our equation in the circuit comes from the well-known tunnelling magnetoresistance (TMR) expression (see the Supporting Information section \secref{section circuit model supp} for derivation) \cite{TMR_expression_2002}:

\begin{equation}
R_{inter}(B) = 
\begin{cases}
  \left( R_P \frac{1 + P^2}{1 + P^2\left[ 2\left( \frac{B}{B_{sw}}\right)^2 -1  \right] } || R_{aux} \right) & \text{if } |B| \leq B_{sw}   \\ \\
   \left( R_P || R_{aux} \right) & \text{if } |B| \geq B_{sw}
\end{cases}
\label{Interlayer resistance}
\end{equation}
\\

\noindent Here, $||$ denotes parallel addition of resistances. $R_P$ is the resistance in the FM state, $P$ is the spin polarization of the layers, $B_{sw}$ is the spin-switching field, and $R_{aux}$ is a parallel resistance representing the non-spin-conserving vertical conduction channels to avoid the $B \to 0$ divergence in the TMR model at $P=1$. Equation~\eqref{Interlayer resistance} thus yields an interlayer resistance with a sharp peak centred at 0~T, saturating near $B_{sw}$ to a value close to $R_P$. $D$ modifies both $R_P$ and $R_{aux}$. 

The use of this TMR-like expression should be understood as an effective description of spin-dependent interlayer transport, rather than as a literal ferromagnet--insulator--ferromagnet junction model. In CrSBr, the adjacent layers are exchange-coupled within a continuous crystal, but the AFM alignment still imposes a spin-dependent constraint on interlayer motion that is relaxed when the layers are field-aligned. The circuit model captures this effect through an effective interlayer resistance. For simplicity, all interlayer links are assigned the same resistance at a given value of $D$; the layer-dependent Stark shift is therefore not treated as a separate microscopic resistance for each interface, but is absorbed into the effective $D$-dependent value of $R_{\mathrm{inter}}$.

Intralayer resistances $R_{\mathrm{intra}}$ are governed by magnon--electron scattering. As the magnetic field increases, the magnon population rises, enhancing the scattering and thus $R_{\mathrm{intra}}$, which accounts for the PMR contribution of the circuit. Experimental MRR measurements on monolayer CrSBr \cite{telford2022coupling} also reveal a gate-dependent NMR component. Accordingly, we model the intralayer resistance as eq. \eqref{Rintra} (Supporting Information section \secref{supplementary: intralayer resistances}).
\begin{align}
R_{intra}(B) = R_0  + R_1 N_{magnons}(B) + R_{NMR}(B)
    \label{Rintra}
\end{align}
Where $R_0$ represents all field-independent contributions to the intralayer resistance, and $R_1$ is a fitting parameter with resistance units, $N_{\mathrm{magnons}}(B)$ is the magnon population at low fields (Supporting Information sections \secref{section circuit model supp} and \secref{fitting PMR supplementary section}), and $R_{\mathrm{NMR}}(B)$ represents the additional contribution responsible for the NMR component (\secref{supplementary: intralayer resistances}).

The term $R_{\mathrm{NMR}}(B)$ should not be interpreted as an additional interlayer spin-filtering contribution. It is included only as a small phenomenological intralayer correction, motivated by the weak low-field NMR reported in monolayer CrSBr \cite{telford2022coupling}, where interlayer spin filtering is absent. In the present multilayer device, the dominant NMR and its strong displacement-field dependence are instead captured by $R_{\mathrm{inter}}(B)$.

Although magnon--electron scattering depends in principle on a transport-weighted
rate rather than on the bare magnon count, we verify that the total population
$N_{\mathrm{magnons}}(B)$ is the quantity that controls the field dependence of
$R_{\mathrm{intra}}$. In the supplementary information, we compute the
magnon--electron conductivity from the linearised Boltzmann equation, using
Fermi-Golden-Rule scattering rates derived from the $s$--$d$ exchange Hamiltonian
that fully account for band velocities and momentum transfer near the Fermi
surface (sections \secref{sec:magnon_electron_model}, \secref{sec:implementation} and \secref{sec:results}). The resulting magnon-electron resistivity follows the same functional form as
$N_{\mathrm{magnons}}(B)$ (figs. \ref{fig:result_Nmag_rho_B}, \ref{fig:result_Nmag_rho_B_Hb} and \ref{fig:result_Nmag_rho_B_Hpa}).

The displacement field is incorporated by fitting the effective
circuit parameters at each experimentally applied value of $D$, rather than by
imposing an analytical law for $R_P(D)$ or $R_{\mathrm{aux}}(D)$. The parameters that are fitted for each $D$ value are shown in table \ref{table: parameter meanings}. The parameters associated with the magnetic response are fixed across the different displacement fields. 

We tuned the 10-layer circuit model using physically realistic parameters to reproduce the results in Figure~\ref{Figures_CrSBr_6}. The fit shows excellent agreement with the experimental MRR traces at different $D$ values (Figure~\ref{Figures_CrSBr_6}b). The
corresponding values are reported in tables \ref{table: fitting parameters Ipa Hpc} and \ref{table: fitting parameters Ipb Hpc}. A key outcome of the model is the evolution of $R_{\mathrm{inter}}$ and $R_{\mathrm{intra}}$ with magnetic field $H$ (Figure~\ref{Figures_CrSBr_6}c).

For all values of $D$, the interlayer resistance decreases towards zero ($-100\%$ MRR) with similar dynamics. The intralayer resistance shows a slight dependence on the electric field \cite{telford2022coupling}. For non-negative $D$, the total MRR is dominated by PMR (magnon--electron scattering), with only a minimal NMR contribution at low magnetic fields, and the intralayer MRR saturates slightly above $40\%$ (Figure~\ref{Figures_CrSBr_6}c). In contrast, a negative $D$ enhances the NMR contribution, as seen for $D=-0.7$ V/nm, where the intralayer MRR saturates just below $40\%$.

The extracted trend shows that negative $D$ reduces the effective interlayer-to-intralayer resistance ratio (Figs.~\ref{fitting PMR}c and \ref{fig:fit_ratio_HpcIpb}a), which favours current passing through a larger number of layers and thus a larger NMR contribution. The layer-resolved current computed from the circuit for positive and negative $D$ illustrates this point (Figs.~\ref{Figures_CrSBr_6}d and \ref{fig:currents_HpcIpb}): for positive $D$ the current is confined to the two bottom layers, whereas for negative $D$ it spreads across the entire stack. The shallow NMR observed at low magnetic fields, including at $D=0$, corresponds to the regime where spin filtering is already partially relaxed while the magnon-driven PMR has not yet become dominant.

    Finally, the same circuit picture also explains transport along the orthogonal in-plane direction, ($I\parallel b$). As shown in the Supporting Information section \secref{sec:circuit_HpcIpb} and figs. \ref{fig:fit_ratio_HpcIpb} and \ref{fig:currents_HpcIpb}, the same two-resistor ladder model reproduces the measured magnetoresistance over the full magnetic-field and displacement-field ranges without introducing a different transport mechanism. The important distinction is the relative contribution of the two channels. For ($I\parallel b$), the resistance change associated with the field-driven reduction of the interlayer contribution, ($R_{inter}$), outweighs the competing change in the intralayer contribution, ($R_{intra}$), at every $D$. The resulting magnetoresistance, therefore, remains negative throughout the entire displacement-field range. By contrast, for ($I\parallel a$), electrostatic gating changes the balance between $R_{intra}$ and $R_{inter}$ sufficiently to reverse the sign of the magnetoresistance. Thus, the contrast between the sign-changing response along the $ a$-axis and the uniformly negative response along the $ b$-axis does not imply different microscopic mechanisms. Instead, it reflects direction-dependent weighting of the same two processes: intralayer magnon scattering and field-induced interlayer spin alignment. The ability of a single circuit model to capture both behaviours, therefore, provides strong support for our central interpretation of the magnetoresistance in CrSBr.

In summary, the circuit model reproduces the measured trends and provides a model-based estimate of the relative contributions of spin filtering and magnon scattering to the magnetoresistance. Within this description, one mechanism or the other dominates depending on whether the current is confined to the bottom layers or distributed across the stack. The fitted interlayer term corresponds to nearly complete spin filtering (Figure~\ref{Figures_CrSBr_6}c), which can be compared with vertical magnetotransport measurements \cite{lin2024influence,liu2025spin,samanta2026electrical}.

\section{Discussion}\label{sec6}

Previous work on CrSBr in a FET structure has revealed several pieces of the puzzle underlying the physics of this material. Our work completes this picture and provides a physical framework that encompasses the contributions of these previous studies in addition to our own. 

The MRR signature and its gate dependence reported in \cite{telford2022coupling}  for different layer thicknesses are fully captured by our circuit model. This is also the case for the monolayer behaviour reported by the same authors, which is represented in our circuit by the NMR element in the layer resistance. A possible physical origin of this contribution was attributed to the existence of a defect band. This effect, incorporated into our circuit model, is also observed in our experimental data when the role of doping in modulating the MRR is decoupled (Figure~\ref{MRR with n and D}c).

Furthermore, the work in \cite{chou2025large} distinguishes between the low- and high-doping regimes of CrSBr FETs, with the MRR behaviour in the high-doping regime also being fully consistent with our experimental and theoretical model. Interestingly, the authors identified a distinct low-doping regime in which the band-gap reduction occurring upon the transition from the AFM to the FM configuration, which is also captured by our DFT calculations, is amplified. As our study has focused on the high-doping regime, our theoretical framework complements this part of the picture by highlighting effects beyond band-gap modulation. Future studies may therefore explore this specific low-doping regime. Furthermore, we have specifically investigated the bottom-contact configuration, which was also used in the aforementioned previous works \cite{telford2022coupling,chou2025large}. By examining spin transport at the layer level in a multilayer system, our circuit model reveals the importance of spin-traffic phenomena and device geometry.

Interestingly, a recent work \cite{wu2022quasi} explored another architecture based on side contacts and revealed an MRR sign opposite to that observed in previous studies, including ours, which used bottom-contact geometries. This finding further highlights the importance of understanding spin traffic and developing a spin-circuit model adapted to the relevant device geometry. Extending our model to the side-contact geometry should provide a route towards explaining this apparent discrepancy or the specific nature of these results.

This framework underscores the importance of contact geometry: in our device, the bottom‐layer contacts govern the displacement‐field response, but alternative designs with side contacts (to all CrSBr layers) or asymmetric source/drain contacts (e.g., top source and bottom drain) could further elucidate these effects.

We have combined ab initio calculations with tri-axial magnetotransport measurements of a bidirectional CrSBr channel in a dual-gate field-effect transistor. Guided by first-principles modelling, our systematic study identifies magnetic-field modulation of the magnon band gap as the origin of the anisotropic positive magnetoresistance component. The double-gate geometry enabled disentangling the effects of carrier density and displacement field, revealing that the latter predominantly controls both the sign and magnitude of magnetoresistance. DFT calculations further show that displacement fields lift layer and spin degeneracies, offering microscopic insight into the observed MRR behaviour. We show that negative MRR arises predominantly from interlayer spin filtering, whereas positive MRR is governed by an intralayer mechanism, and we quantify their anisotropic contributions. 
First-principles magnon-electron scattering calculations quantitatively reproduce the positive magnetoresistance, capturing its magnetic-field, temperature, and anisotropy dependence, and identify the field-driven closure of the spin-wave gap as its microscopic origin. By independently mapping carrier density and displacement field across the full gate space, we further establish that the displacement field, rather than doping, controls the sign and amplitude of the magnetoresistance for every field orientation.

Finally, we developed a spin-circuit model that incorporates both microscopic mechanisms and device geometry. By fitting the model to the experiment, we quantify the relative contributions of the underlying mechanisms and demonstrate how spin-dependent current distributes across layers under applied electric and magnetic fields. The same ladder model, without modification, reproduces the uniformly negative response measured along the high-mobility b-axis, confirming that the contrasting sign behaviour along the two current directions reflects a direction-dependent weighting of the same two microscopic processes rather than distinct mechanisms. Beyond our own device, this framework encompasses and reconciles previously reported gate- and thickness-dependent magnetoresistance signatures in CrSBr field-effect transistors, unifying these observations within a single spin-traffic picture.

These findings provide a blueprint for spintronic circuit design using antiferromagnetic semiconductors. Engineering the electric potential landscape with multi-gate FET architectures emerges as a promising strategy to control spin-current distribution and to realize advanced spin-logic devices.

\section{Methods}\label{sec7}
\noindent\underline{Sample fabrication}\\
All exfoliation and assembly steps were carried out in an argon-filled glovebox (O$_2$ and H$_2$O levels $<$ 1 ppm). The electrodes were patterned on a Si/SiO$_2$ substrate via electron-beam lithography. All 2D materials were then exfoliated in the glovebox onto a PDMS stamp. A layer-by-layer pickup procedure using a PC stamp was used to assemble the stack, which was subsequently transferred onto the pre-patterned electrodes and encapsulated with an hBN capping layer.

\medskip
\noindent\underline{Magnetotransport}\\
All electrical measurements were performed in a Quantum Design PPMS DynaCool cryostat, at temperatures between 2 and 300~K and in magnetic fields of up to 9~T. The channel resistance and its dependence on the top- and bottom-gate voltages were measured with Keithley 2614B source-measure units. The samples showed no breakdown or degradation over the whole measurement campaign, which consisted of continuous 24~h sessions over several weeks.

\medskip
\noindent\underline{Modelling}\\
For the DFT calculations, we employed the electronic-structure package SIESTA \cite{Soler2002_primary_SIESTA_paper}. The exchange--correlation energy was described within the generalized gradient approximation (GGA) using the PBEsol functional \cite{PBEsol_2008}. Standard solid-state pseudopotentials were taken from the PseudoDojo database \cite{PseudoDojo_2018,PseudoDojo_PSML_2018}. Well-converged results were obtained with a mesh cutoff of 1200 Ry for bulk CrSBr and 4000 Ry for the six-layer system. The Brillouin zone was sampled with a $12 \times 12 \times 12$ Monkhorst--Pack mesh \cite{Monkhorst_1976} for bulk and $12 \times 12 \times 1$ for the six-layer system. To account for strong electron correlations, we adopted a DFT+$U$ approach using the rotationally invariant Liechtenstein formulation \cite{Monteseguro_2023_DFT+U_SIESTA,Liechtenstein_dft+U_rotational_invariant}, with $U=4$ eV for the on-site Coulomb repulsion and $J=1$ eV for the on-site exchange interaction. Spin--orbit coupling was included via fully relativistic pseudopotentials \cite{Cuadrado2012_SOC}, and van der Waals interactions were treated with Grimme-D3 semi-empirical corrections \cite{DFT_D3_2010,DFT_D3_2011}. Further post-processing was carried out with SISL visualization tools \cite{zerothi_sisl}. Exchange interactions were computed via Green's functions using the TB2J package \cite{TB2J_2021}, and magnon dispersions were obtained with SpinW \cite{Toth_2015} using Colpa's method \cite{colpa1978diagonalization}.

The magnon density of states (MDOS) was calculated as
\begin{equation}
m\text{-DOS}(E) = \frac{1}{N} \sum_{i=1}^N \delta(E - E_i),
\label{MDOS}
\end{equation}
where $N$ is the total number of eigenenergies, $E$ is the energy, $E_i$ is the $i$-th eigenenergy, and $\delta$ is the Dirac delta function. Here, the $E_i$ values correspond to the magnon eigenvalues obtained with SpinW.  A Gaussian broadening of 0.02 meV was applied to the Dirac deltas for a smooth spectrum.

The magnetoresistance ratio (MRR) was computed as
\begin{equation}
\mathrm{MRR}(B,T) = \frac{R(B,T) - R(0,T)}{R(0,T)} ,
\label{MRR}
\end{equation}
where $R(B,T)$ is the resistance at field $B$ and temperature $T$.

\section{Data Availability}
The data that support the findings of this study are available within the paper and its Supplementary Information. The raw magnetotransport data and the input and output files of the DFT, spin-wave and circuit-model calculations are available from the corresponding authors upon request.

\section{Author Contributions}

J.-F.D. and M.S. conceived the project and experiment and proposed the initial physical interpretation, which served as the starting point for the theoretical development.

M.S. fabricated the samples and led the measurements, which he and D.B. carried out together in close collaboration. N.K.R. contributed to the experimental measurements and to the development of the magnetotransport experimental setup. A.G. developed the motorized transfer station. N.C. performed the micro-Raman measurements, which A.O. supervised and analysed. J.R. and B.D. provided technical and scientific support, respectively, on the cryogenic systems. T.T. and K.W. provided the hBN crystals.

J.G.A. developed the theoretical calculations, simulations, and modelling, in close discussion on the physical aspects with J.-F.D., M.S., J.H.G., and D.L.E. D.L.E. supervised the theoretical development, calculations, and simulations. D.L.E., J.H.G., and S.R. contributed to the data analysis and to the supervision of the theoretical work.

J.-F.D. coordinated the scientific collaboration, together with D.L.E. for the theory. The manuscript writing was led by M.S. and J.G.A., together with J.-F.D. and D.L.E. All authors contributed to the broad-scope scientific discussion, to the editing and reviewing of the manuscript, and to the collaborative multidisciplinary discussion.

\textcolor{black}{
\section{Acknowledgements}\label{sec8}
We thank Fabien Chevrier for technical assistance and the STnano nanofabrication platform. We acknowledge Paul Noel and Manuel Bibes for scientific discussions. We also acknowledge the computer resources at MareNostrum and the technical support provided by the Barcelona Supercomputing Center (RES-FI-2025-3-0058).
}
\textcolor{black}{
\section{Funding}\label{sec9}
J.F.D. acknowledges the Agence Nationale de la Recherche for financial support through the grants MixDFerro (ANR-21-CE09-0029) and SOFIANE (ANR-23-CE09-0007), and IdEx Unistra (ANR 10 IDEX 0002), SFRI STRAT'US project (ANR 20 SFRI 0012), and EUR (QMat-ANR-18-EUR-0016) under the framework of the French Investments for the Future Program. This work was supported by a government grant operated by the French National Research Agency as part of the France 2030 program PEPR Spin, reference ANR-24-EXSP-0008 (MULTISPIN). M.S. and J.F.D. acknowledge Eucor - The European Campus for financial support through the QUSTEC funding from the European Union's Horizon 2020 research and innovation program under the Marie Sklodowska-Curie Grant Agreement No. 847471 and support of the Institut Universitaire de France (IUF). K.W. and T.T. acknowledge support from the JSPS KAKENHI (Grant Numbers 21H05233 and 23H02052), the CREST (JPMJCR24A5), JST and World Premier International Research Center Initiative (WPI), MEXT, Japan.
}
\textcolor{black}{
J. G. A. acknowledges financial support from the MICIU under grant no. PRE2022-103460, funded by CEX2021-001214-S-20-8/MCIU/AEI/10.13039/501100011033 and by FSE+. D. L. E. acknowledges financial support from the MICIU and the European Union through the NextGenerationEU grant PRTR-C17.I1. J. H. G. acknowledges funding from the European Union (ERC, AI4SPIN, grant agreement no. 101078370). J. G. A., D. L. E., J. H. G., and S. R. acknowledge support from PID2019-106684GB-I00, funded by MCIN/AEI/10.13039/501100011033 and FEDER, UE, as well as from PID2022-138283NB-I00, funded by MCIU/AEI/10.13039/501100011033. They also acknowledge support from the Generalitat de Catalunya (SGR) and from the European Union ``NextGenerationEU/PRTR'' under grant PCI2021-122035-2A-2a, as well as funding from the European Union's Horizon 2020 research and innovation programme under grant agreement no. 881603. ICN2 is funded by the CERCA Programme/Generalitat de Catalunya and supported by the Severo Ochoa Centres of Excellence programme (grant CEX2021-001214-S).
}
\section{Competing Interests}
The authors declare no competing interests.

\section{Ethics Declaration}
Ethics declaration: not applicable.

\end{document}